\documentclass[amssymb,prl,aps,twocolumn,amsmath,showpacs]{revtex4} 
\usepackage[T1]{fontenc}
\usepackage{graphicx}
\usepackage{color}
\usepackage{amsmath}
\usepackage{ulem}

\newcommand{\Edrift}{E^{\rm{drift}}}
\newcommand{\Ediff}{E^{\rm{diff}}}

\newcommand{\kT}{k_{\rm{B}} T}
\newcommand{\tW}{t_{\rm{W}}}

\newcommand{\Escbf}{\bf{E_{\rm{sc}}}}
\newcommand{\Ebf}{\bf{E_0}}
\newcommand{\ybf}{\bf{y}}

\newcommand{\kvec}{\bf{k}}

\newcommand{\tauW}{\tau_{\rm{W}}}
\newcommand{\kL}{k_{\rm{L}}}
\newcommand{\yL}{y_{\rm{L}} }

\newcommand{\kc}{k_{\rm{c}} }
\newcommand{\no}{n_{\rm{o}} }
\newcommand{\nex}{n_{\rm{e}} }

\newcommand{\IL}{I_{\rm{L}}}
\newcommand{\Ip}{I_{\rm{p}}}

\newcommand{\vdrift}{v_0^{\rm{drift}}}
\newcommand{\vdiff}{v_0^{\rm{diff}}}

\newcommand{\Is}{I_{\rm{s}}}
\newcommand{\IW}{I_{\rm{W}}}

\newcommand{\Isat}{I_{\rm{sat}}}

\newcommand{\lR}{l_{\rm{R}}}

\newcommand{\kB}{k_{\rm{B}}}

\newcommand{\Esc}{E_{\rm{sc}}}
\newcommand{\Escinf}{E_{\rm{sc}}^{\rm{\infty}}}

\newcommand{\Itilde}{\tilde{I}}
\newcommand{\gradp}{\nabla^2_{\perp}}

\newcommand{\dnmaxinf}{\delta n _{\rm{max}} ^{\infty}}
\newcommand{\dninf}{\delta n ^{\infty}}
\newcommand{\Deltaninf}{\Delta n ^{\infty}}

\begin{document}
\title{Photorefractive writing and probing of anisotropic linear and non-linear lattices}
\date{\today}

\author{Rapha\"el Allio$^{1,2}$, Diego Guzm\'an-Silva$^1$, Camilo Cantillano$^1$, Luis Morales-Inostroza$^1$, Dany Lopez-Gonzalez$^1$, Sebasti\'an Etcheverry$^{3}$, Rodrigo A. Vicencio$^1$ and Julien Armijo$^{1}$}

\email[Corresponding author: ]{julienarmijo@gmail.com}

\affiliation{$^1$Departamento de Fisica, MSI-Nucleus on Advanced Optics, and Center for Optics and Photonics, Facultad de Ciencias, Universidad de Chile, Santiago, Chile\\
$^2$Universit\'e de Rennes I, France\\
$^3$Center for Optics and Photonics, Universidad de Concepci\'on, Concepci\'on, Chile}

\begin{abstract}
We study experimentally the writing of one- and two-dimensional photorefractive lattices, focusing on the often overlooked transient regime. 
Our measurements agree well with theory, in particular concerning the ratio of the drift to diffusion terms.
We then study the transverse dynamics of coherent waves propagating in the lattices, in a few novel and simple configurations.
For focused linear waves with broad transverse spectrum, we remark that both the intensity distributions in real space ("discrete diffraction") and Fourier space ("Brillouin zone spectroscopy") reflect the Bragg planes and band structure.
For non-linear waves, we observe modulational instability and discrete solitons formation in time domain.
We discuss also the non-ideal effects inherent to the photo-induction technique : anisotropy, residual nonlinearity, diffusive term, non-stationarity.
\end{abstract}

\maketitle

\section{I. Introduction}

The photorefractive effect is the process  by which refractive index changes can be induced in photosensitive crystals as a consequence of illumination with light patterns.
This effect is complex, intrinsically nonlocal and anisotropic, and features various regimes \cite{yeh93, hall85, frejlich07, kukhtarev78, zozulya95}.
In the last decades, the photorefractive effect has been often used for generating waveguide arrays (photonic crystals) and study the linear and non-linear propagation of light waves inside them. Remarkable realizations included the observation of discrete optical solitons \cite{fleischer03b}, discrete optical vortices in 2D lattices \cite{neshev04}, or Anderson localization of light in disordered landscapes \cite{schwartz07}, among many others.

Despite these numerous works, systematic studies of photorefractive lattice writing and wave propagation inside them, providing quantitative comparisons of measurements with theories, are rare.
In this paper, we study the  photorefractive lattice writing process, especially in the often overlooked -but nonetheless relevant- transient regime, and several cases of wave propagation, providing some new observations in simple configurations. 
We compare measurements with theories and in some cases, for the first time to our knowledge, to simulations with only directly calibrated (and non-adjustable) parameters, using a new lattice calibration method \cite{Armijo14}.
We discuss overall the validity of the various approximations, and the strength of non-ideal effects, to improve our understanding of photorefractive lattice experiments.

In section II, we briefly review the standard theory of photorefractive writing and probing.
In section III, we study the transient photorefractive writing, for the simplest case of a 1D lattice. 
We find good agreement  with standard theory concerning the role of various parameters, and the ratio between the drift and diffusion photorefractive terms which is linear in lattice period.
In section IV, we study the propagation of simple linear and non-linear waves in regular lattices. We describe an interesting analogy between the linear patterns of  discrete diffraction (in real space) and Brillouin zone spectroscopy (in Fourier space), the Bragg planes and band structure being apparent on both types of pictures, and we compare both measurements to simulations.
Finally, we study nonlinear effects in some new configurations.
Modulational instability is observed in a quasi-1D geometry at the center of the Brillouin zone for a focusing non-linearity, and is absent for a defocusing one.
The transition from discrete diffraction to a discrete soliton is observed in time, due to the differential writing speed for the lattice and nonlinear effects.

\section{II. Theory of photorefractive lattice writing and probing}

\subsection{1. Photorefractive effect}

Let us first recall the relevant theoretical frame.
The basic mechanism of the photorefractive effect is the photogeneration of mobile charge carriers, generally assumed to be only electrons, which are then subject to displacement in the crystal, purely diffusive (from light to shadow), or driven by an externally applied electric field $\Ebf$. Their recombination at different locations gives rise to a permanent space-charge electric field $\Escbf$, which, via the linear (Pockels) electro-optic effect, creates modulations of the refractive index inside the crystal.

In the particular case of the most often used strontium barium niobate (SBN) crystals, which belong to the point 4 mm symmetry group, and assuming $\Escbf$ oriented along the crystalline axis, transverse direction $y$ (which is valid in the drift dominated case, see below), the extraordinary and ordinary refractive index changes in the crystal can be written (see, e.g., \cite{denz03})
\begin{align}
\delta \nex = \frac{1}{2} \nex^3 r_{33} \Esc, \\
\delta \no = \frac{1}{2} \no^3 r_{13} \Esc,
\label{eq.deltan}
\end{align}
where $\nex$ and $\no$ are respectively the extraordinary and ordinary refractive indexes in zero electric field, and $r_{ij}$ the relevant electro-optic coefficients
\footnote{Here only one component of electric field contributes, this is due to the particular symmetry of SBN crystals \cite{denz03}}.
The standard model used to describe the dynamics of charge carrier generation, displacement, and the resulting field and refractive index modulations, is due to Kukhtarev et al. \cite{kukhtarev78}.
Within this model, most works consider only the steady-state, but the transient regime can also be very relevant experimentally, as we show below.

\subsubsection{Isoptropic approximation}

As a first step, theoretical works have treated purely one-dimensional situations \cite{segev94, christodoulides95}, considering $\Ebf$$ =E_0 \ybf$ and $\Escbf$ $=\Esc \ybf$.
In this frame, neglecting the dynamics, and any photovoltaic contribution (which is valid for SBN crystals), the Kukhtarev model allows to derive the stationary space charge electric field 
\begin{equation}
\Escinf = \frac{1}{1 + \Itilde} \left( E_0 - \frac{\kT}{e} \frac{\partial \Itilde }{\partial y} \right),
\label{eq.Esc}
\end{equation}
were $T$ is the temperature, $\kB$ the Boltzmann constant, $e$ the electron charge, and $\Itilde = \IW /\Isat$ is the writing beam intensity $\IW$ normalized to the "dark intensity" or "saturation intensity" $\Isat$, which is a phenomenological parameter accounting for the probability that electrons are thermally excited in the conduction band.
From Eq. \ref{eq.Esc} one obtains the stationary refractive index change
\begin{equation}
\dninf= \dnmaxinf  \frac{1}{1+ \Itilde},
\label{eq.dninf}
\end{equation}
where we note
\begin{equation}
\dnmaxinf = 0.5 \nex^3 r_{33} \Escinf.
\label{eq.dnmaxinf}
\end{equation}
Note that the sign of the photorefractive effect depends on the sign of $\Esc$, thus, nonlinearities of focusing or defocusing type can be generated.

The simplest case is when the diffusion term can be neglected (which we check in section IV), then the non-local contribution disappears and one simply has 
\begin{equation}
\Esc = \frac{E_0 }{1 +\Itilde}.
\label{eq.Escloc}
\end{equation}

The isotropic approximation consists in assuming that this is valid not only as a scalar expression, but also, vectorially, i.e., that Eq. \ref{eq.Escloc} can be written with the vectorial electric fields $\Escbf$ and $\Ebf$ (see, e.g., \cite{fleischer03b}). 
We are not aware of quantitative studies or verifications of the validity of such approximation.

\subsubsection{Full anisotropic model}

The isotropic model of Eq. \ref{eq.Esc} and \ref{eq.Escloc} is useful in 1D, and to intuitively grasp the interplay between the different physical mechanisms, for example, to compare the importance of diffusive vs drift mechanisms. 
However, in the general case of 2D refractive index landscapes, the intrinsic crystal anisotropy and the electric field $\Ebf$ strongly destroys the isotropy of the system.

To describe the photorefractive effect in the general case, an anisotropic model has been proposed by Zozulya and Anderson \cite{zozulya95}, which has the structure of a nonlinear problem for the electrostatic potential $\phi = \phi_0 -  \vert E_0 \vert y$ such that 
\begin{equation}
\Escbf=-\nabla \phi, 
\label{eq.phi}
\end{equation}
where the light-induced potential $\phi_0$ is separated from  the contribution of the external field $\Ebf $$=E_0 \ybf$.
Starting form the 3D equations of the Kukhtarev model, assuming a slowly varying light intensity field $I$, one obtains, in the stationary regime, the governing equation \cite{zozulya95, denz03} 
\begin{equation}
\begin{split}
\nabla ^2 \phi_0  + \nabla &  \ln (1 +\Itilde ) \nabla \phi_0  = \vert E_0 \vert \frac{\partial \ln (1+\Itilde)}{\partial y} \\
 & \quad \quad  -\frac{\kB T}{e} \left[ \nabla^2  \ln(1+\Itilde) + (\nabla \ln(1+\Itilde))^2  \right],
\end{split}
\label{eq.anis}
\end{equation}
In the right hand side of Eq. \ref{eq.anis}, the first term proportional to $\vert  E _0 \vert$ is the drift term, and the second, proportional to $\kB T/e=D/\mu$ (where $D$ is the diffusion coefficient and $\mu$ the electron mobility), the diffusive term.

If one neglects the diffusive effect (which we discuss in Section IV), as, e.g., in \cite{desyatnikov06, terhalle07, zhang07}, Eq. \ref{eq.anis} becomes
\begin{equation}
\nabla ^2 \phi_0  + \nabla   \ln (1 +\Itilde ) \nabla \phi_0  = \vert E_0 \vert \frac{\partial \ln (1+\Itilde)}{\partial y}. 
\label{eq.anis2}
\end{equation}

\subsubsection{Time evolution of the refractive index}

The temporal evolution of the photorefractive effect can have primary importance in experiments.
In general, photorefractive recording follows a damped oscillatory behavior  \cite{hall85, frejlich07}, however for small index changes, and/or when the writing beam intensity $\IW$ is so low that the associated timescale is much longer than the material's intrinsic microscopic timescales, the writing (and also erasing) processes are well described by an exponential dependence with writing time $\tW$ as
\begin{equation}
\delta n= \delta n^{\infty} [1 - \exp (-\tW / \tauW)].
\label{eq.dnexp}
\end{equation}
For our parameters, the time constant $\tauW$ is determined by the rate of carrier generation set by $\IW$ since all microscopic timescales are much shorter, thus we have \cite{hall85}
\begin{equation}
\frac{1}{\tauW} \propto \IW + \Isat,
\label{eq.tauW}
\end{equation}
and the exact value of $\tauW$ depends on several parameters including the lattice period. For our parameters typically $\tauW \sim 10-100$s for linear lattice writing, while the development of nonlinear patterns is typically one order of magnitude slower (see Section IV.2).

\subsection{2. Wave propagation in a photo-written lattice}

\paragraph{Linear case :}
Considering now (regardless of its origin) a transverse refractive index landscape $\delta n (x,y)$, invariant in $z$, the propagation of a wave of amplitude $\Psi(x,y,z)$ and vacuum wavelength $\lambda=2\pi/k_0$ in the paraxial approximation obeys a transverse (2+1)D Schr\"odinger equation \cite{kivshar03}
\begin{equation}
i \frac{\partial \Psi}{\partial z} = -\frac{1}{2 \beta_0} \gradp \Psi - \frac{\beta_0}{n_0} \delta n (x, y) \Psi ,
\label{eq.se}
\end{equation}
where $n_0=\nex$, $\beta_0=2\pi n_0/\lambda$ is the propagation constant in the crystal, $\gradp = \left(\frac{\partial ^2}{\partial x ^2} + \frac{\partial ^2}{\partial y ^2} \right)$ denotes the transverse laplacian operator, the longitudinal (propagation) coordinate $z \leftrightarrow t $ plays the role of the time $t$, and the potential $V(x,y)$ is here replaced by the refractive index, i.e. $V(x,y) \leftrightarrow - \delta n(x,y)$. The correspondence to the Schr\"odinger equation is complete with the additional replacement of the particle mass by the refractive index $m \leftrightarrow n_0$ and the reduced Planck constant $h/2 \pi \leftrightarrow \lambda/2\pi$.

\paragraph{Non-linear case :} 
For more intense beams, photoexcitation of carriers by the probe beam does influence the refractive index pattern, i.e., non-linear propagation occurs, and the stationary refractive index is also a function of the beam intensity $I$.
In the general anisotropic case, one has to solve the system of Eq. \ref{eq.se} combined with Eq. \ref{eq.phi} and Eq. \ref{eq.anis} (or Eq. \ref{eq.anis2} if one neglects the diffusive contribution). This is done in several works (see, e.g., \cite{desyatnikov06, zhang09}), and it allows to reach a fairly good agreement between simulations and measurements.

Sometimes one also finds the more simplified isotropic approximation, also neglecting the diffusive term, which consists in simply using Eq. \ref{eq.Escloc} for the nonlinear problem. 
Then, the propagation of a probe beam in the crystal is approximated by a (2+1)D nonlinear Schr\"odinger equation (NLSE) with saturable nonlinearity in the form 
\begin{equation}
i \frac{\partial \Psi}{\partial z} = -\frac{1}{2 \beta_0}\gradp \Psi - \frac{\beta_0}{n_0} \delta n (x, y)\Psi - \Gamma \frac{ \Psi}{1+ \vert \Psi \vert^2/\Isat},
\label{eq.nlse}
\end{equation}
where $\Gamma=(1/2) k_0^2 \nex^3 r_{33} E_0 $ is the effective non-linear coefficient.

To get an idea of the importance of the photorefractive anisotropy in typical experiments, one can observe, in \cite{zozulya95}, the simulations using the full anisotropic model of Eq. \ref{eq.anis}, including the diffusive term, for a gaussian beam propagating with attractive nonlinearity.
In Fig. 4 and 7 of \cite{zozulya95}, one notes several differences in the refractive index profiles in the two transverse directions ($y$ and $x$ in our notations). In particular, the profiles in the c-axis direction display not only a local minimum, but also two local maxima aside of it, which has been observed experimentally in \cite{petter02}.
More complex situations have also been studied, for example the possibility to obtain a hybrid (focusing and defocusing) nonlinearity \cite{zhang07, zhang09}.
Another consequence of the particularity of the photorefractive nonlinearity, is that even the formation of a continuous 2D soliton is not trivial, requiring specific parameter ranges, and has given rise to debate \cite{duree93, zozulya96}.

\section{III. Temporal study of photorefractive lattice writing}

\subsection{1. Experimental set-up}

\begin{figure}
\includegraphics[width=8.5cm]{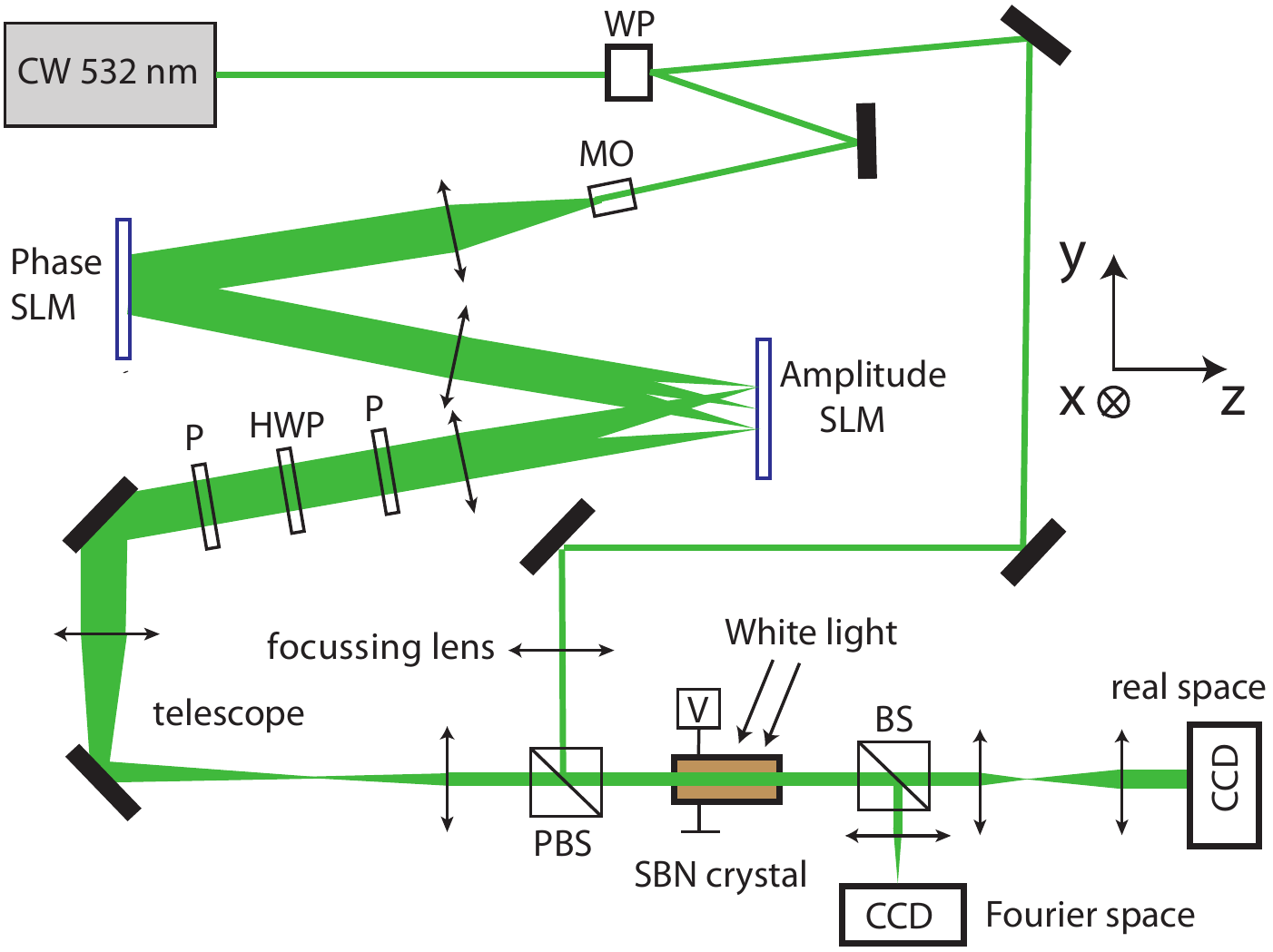}
\caption{Experimental set-up. WP : Wollaston prism. MO : microscope objective. P : polarizer. HWP : half-wave plate. PBS : polarizing beam splitter. BS : beam splitter.}
\label{fig.setup}
\end{figure}

For inducing and studying photorefractive lattices, we use standard techniques, as sketched in Fig.\ref{fig.setup}. A cw laser beam at wavelength $\lambda=532$nm is split in two components of polarization. The ordinary polarized beam is used as a lattice writing beam, modulated in real space with a phase SLM (Holoeye Pluto) and dynamically filtered in Fourier space using an amplitude SLM (Holoeye LCR-1080). 
This configuration allows us to realize clean non-diffracting lattice beams in any 2D geometry, provided that the transverse spectrum of the lattice waves is contained in a circle \cite{boguslawski11}.
On the other hand, the extraordinary polarized beam is used as a probe beam, which eventually is shapen anisotropically using a cylindrical lens, or focused with regular lenses. 
We use a $10\times5\times2$ mm$^3$, 0.005\% CeO2 doped SBN:75 crystal, whose relevant electro-optic coefficients are $r_{33} = 1340$pm/V and $r_{13} = 67$pm/V \footnote{In our notation, the c-axis of the crystal is $y$.}.
We apply no background illumination during writing and, from the erasing time of lattice patterns in the dark ($\sim$1 day), we estimate the saturation (dark) intensity in our crystals $\Isat\sim1\mu$W/cm$^2$ i.e., we work at high saturation $\IW \gg \Isat$.

\subsection{2. Writing efficiency in 1D lattices}

\begin{figure*}
\includegraphics[width=16cm]{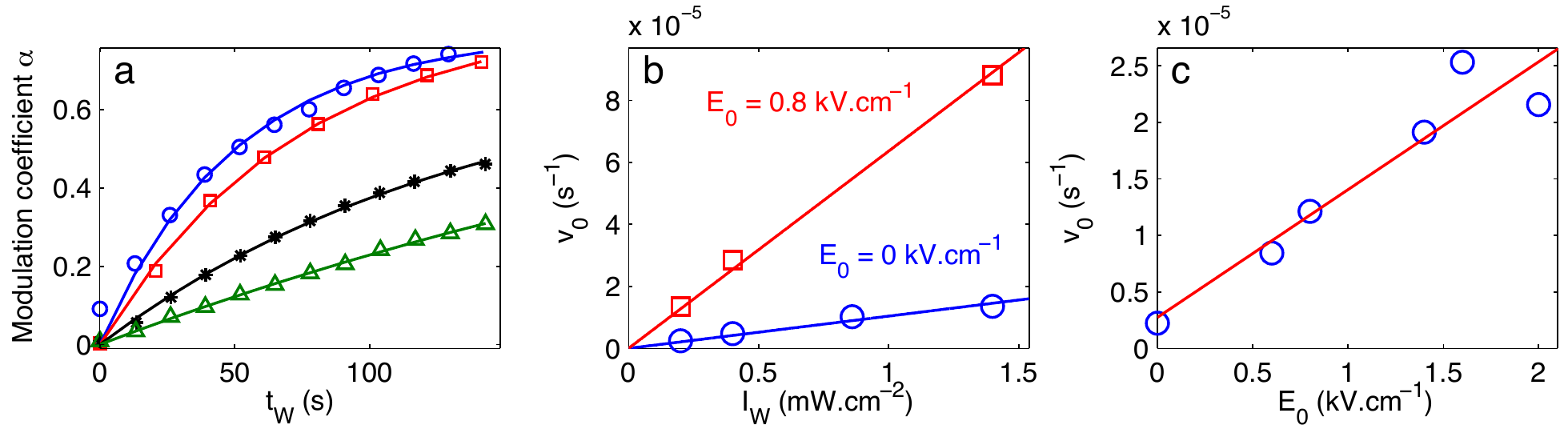}
\caption{Influence of external parameters on writing efficiency for a 1D lattice of period $d=10\mu$m. (a) Determination of initial writing speed $v_0$ from fits of the modulation coefficient $\alpha$ as function of writing time $\tW$ with Eq. \ref{eq.dnexp}, for $E_0=0$V, and average writing beam intensities $\IW=0.2, 0.4, 0.86, 1.4$mW/cm$^2$ (from bottom to top).
(b) Initial writing speed $v_0$ vs $\IW$ for $E_0=0$ and $0.8$kV/cm. Solid lines show linear fits according to Eq. \ref{eq.v0}.
(c) Initial writing speed $v_0$ vs applied field $E_0$, for $\IW=0.2$mW/cm$^2$. The solid line is a linear fit to the data.}
\label{fig.1dwriting}
\end{figure*}

In this section we present time-resolved measurements of the writing process for regular lattices, using a new calibration technique presented in \cite{Armijo14}. 
For simplicity, we treat only 1D lattices oriented in the strong, c-axis direction $y$. 
As noted in \cite{Armijo14}, 2D lattices involves a higher degree of non-ideal effects, thus, 1D lattices are more favorable for carrying basic quantitative studies.

The experimental sequence consists in first writing a lattice during a writing time $\tW$ with a writing beams of average intensity $\IW$, and a bias field $E_0$.
Noting the intensity-dependent refractive index modulation (the refractive index change minus its value at zero intensity)
\begin{equation}
\Delta n (I) = \delta n (I) - \delta n (0),
\label{eq.deltanMod}
\end{equation}
as in \cite{Armijo14}, we assume $\Delta n (I)$ proportional to the lattice intensity, i.e., of the form \footnote{It is reasonable to assume a constant refractive index along $z$ since the absorption coefficient for our crystals is $\alpha<0.4/$cm (source : Altechna).
Assuming a sinusoidal refractive index profile may not seem valid because we work at high saturation $\IW \gg \Isat$ so that distortions are expected in the stationary regime. However, we work in the transient regime, where the writing speed is proportional to the intensity (see Eq. \ref{eq.tauW}), thus the local refractive index in is expected to be proportional to $\IW+\Isat\simeq \IW$. Our sinusoidal assumption also neglects any parasitic nonlinear effects for the writing beam.}
\begin{equation}
\Delta n (y)= \Delta n_0  \sin^2(\kL y/2 + \phi_L),
\label{eq.deltan}
\end{equation}
where $\kL=\pi/d$ and $d$ is the lattice period.

In a second step, we shut off the lattice waves and send a probing plane wave at very low intensity into the crystal.
The intensity distributions at the crystal output face are recorded on the real space CCD camera, and integrated in the $x$ direction so that we consider only profiles $I(y)$.
To quantify the strength of the lattices, we use a fitting function, as in  \cite{Armijo14}, of the form 
\begin{equation}
I(y) = I_0 \big[1 + \alpha \cos(\kL y  + \phi_1) \big],
\label{eq.fit}
\end{equation}
where $\alpha$ is a modulation coefficient.

Analytic theories for the photorefractive effect are available in the steady-state regime (see Section II).
However, observing the convergence and stabilization to a steady-state is not always easy to achieve in experiments, since we often observe parasitic effects and instabilities, especially for 2D lattices (see, e.g., the 2D lattices calibrations in \cite{Armijo14}).
On the other hand, it is much easier to extract meaningful information from the behavior at short times, where the refractive index is not yet strong and non-ideal effects are weaker.
Last, but not least, the transient photorefractive regime gives access to adjustable lattice strength \cite{boguslawski13, Armijo14}.
Thus, we consider for our study the initial rate of refractive index change
\begin{equation}
v_0= \left( \frac{\partial \Delta n_0}{ \partial \tW} \right)_{\tW=0}.
\label{eq.v0}
\end{equation}
To estimate $v_0$ from measured data, we plot the modulation ratio $\alpha$ as function of $t_W$ and fit its initial behavior with an exponential function
\begin{equation}
\alpha= \alpha_\infty [1 - \exp (-\tW / \tauW)].
\label{eq.aexp}
\end{equation}
Fig. \ref{fig.1dwriting}.a shows some examples of this procedure, for $E_0=0$, $d=10\mu$m and different writing beam intensities $\IW$.
Our method for absolute calibration of $\Delta n_0$ \cite{Armijo14} then allows to convert the initial slope $\partial \alpha / \partial \tW$ into $v_0$. For moderate lattice strengths $\Delta n_0$ (which is often valid at short times), $\Delta n_0$ is simply proportional to $\alpha$.

In Fig. \ref{fig.1dwriting}.b, we show $v_0$ as function of $\IW$, for $d=10\mu$m, and for a drift-dominated regime ($E_0=0.8$kV/cm, squares), and a diffusion-dominated regime ($E_0=0$, circles).
From Eq. \ref{eq.deltanMod}, we have $ \Deltaninf = - \dnmaxinf \IW/ (\IW + \Isat)$, and using Eq. \ref{eq.tauW} and $\IW \gg \Isat $ one sees that the initial writing speed is expected to be
\begin{equation}
v_0= \frac{\Deltaninf}{\tauW }\propto \dnmaxinf \IW,
\label{eq.v0}
\end{equation}
i.e. simply linear in $\IW$. 
The data in Fig. \ref{fig.1dwriting}.b are in very good agreement with fits using Eq. \ref{eq.v0} (solid lines). The ratio of the fitted slopes is 6.1, in good agreement with the theoretical expectation of 4.9 using Eq. \ref{eq.ratio}.

In Fig. \ref{fig.1dwriting}.c, we plot $v_0$ for a lattice of period $d=10\mu$m as function of $E_0$.
The dependence is nearly linear, which is in agreement with the expectation from the isotropic model (Eq. \ref{eq.Esc}).
The non-zero offset at $E_0 =0$ is due to the diffusive term of the photorefractive effect.
For our typical working value $E_0 = 1.5$kV/cm, this offset is small and thus we reasonably consider that the photorefractive effect lies in the drift-dominated regime.

\subsection{3. Drift \textit{vs} diffusive photorefractive effects}

Let us quantify the relative importance of the drift vs diffusion terms.
For fixed lattice period $d$ and writing intensity $\IW$, according to Eq. \ref{eq.v0}, we expect that the initial writing speed is simply proportional to $\dnmaxinf$ given by Eqs. \ref{eq.Esc} and \ref{eq.dnmaxinf}.
Writing the modulation of electric field $\Delta E = \Esc(I) -E_0=\Delta \Edrift + \Delta \Ediff $ as
\begin{equation}
\Delta E = \frac{1}{1+ \Itilde} \left( - E_0 \Itilde - \frac{\kT}{e} \frac{\partial \Itilde}{\partial y}  \right),
\label{eq.deltaEsc}
\end{equation}
one obtains, using Eq. \ref{eq.fit} and considering only the writing velocity for the lattice maxima,
\begin{equation}
\frac{\vdrift}{\vdiff}=\frac {\Delta \Edrift}{\Delta \Ediff} = \frac{e}{\kT} \frac{E_0}{k_L}.
\label{eq.ratio}
\end{equation}
In particular one can express the electric field for which the diffusive and drift terms have the same strength \cite{hall85} as $E_{\rm{D}}= \left( \kT /e\right) \kL$.

\begin{figure}
\includegraphics[width=6cm]{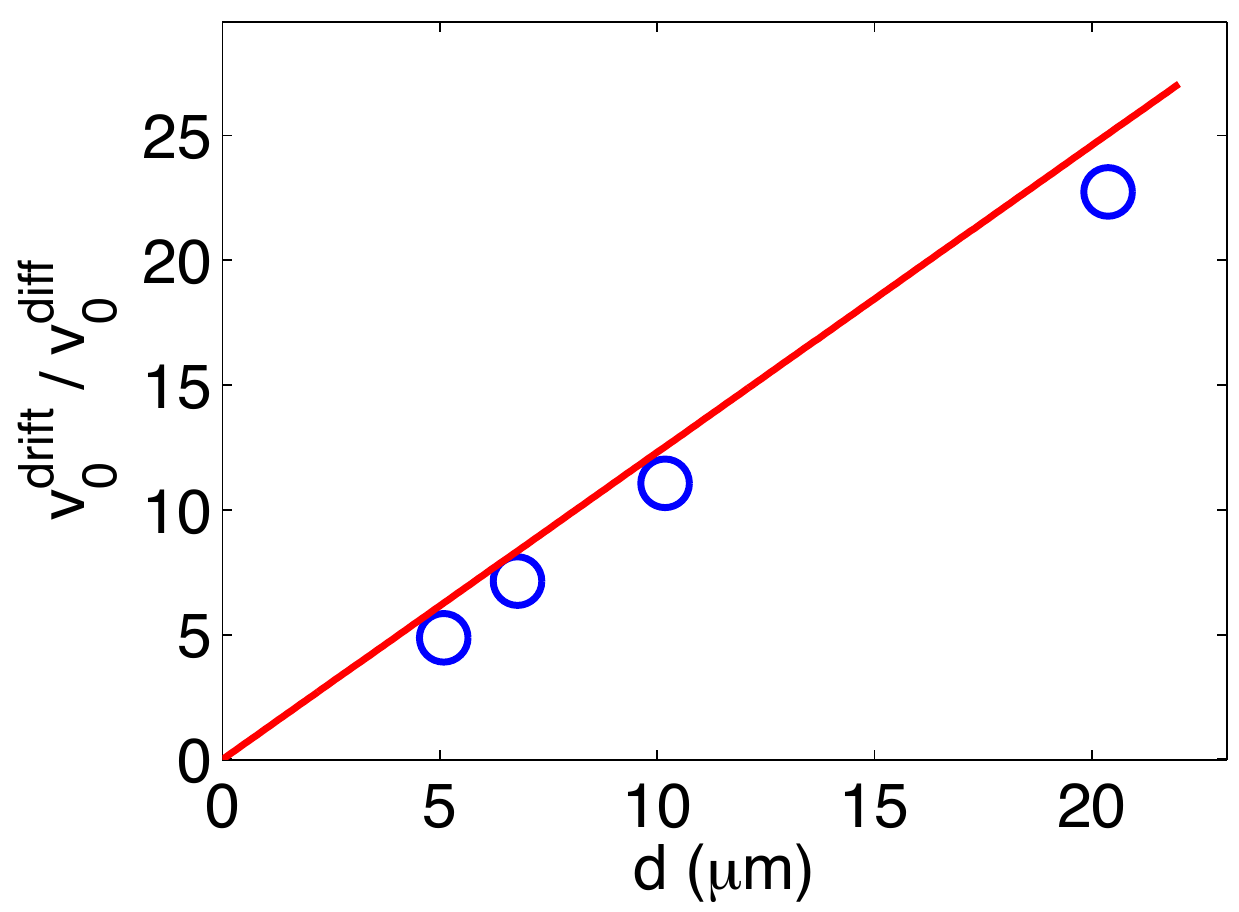}
\caption{Ratio of initial writing speeds for drift ($E_0= 2 $kV/cm) and diffusive ($E_0=0$) photorefractive mechanisms versus lattice period $d$, for $\IW=1$mW/cm$^2$. The solid line is the theoretical prediction Eq. \ref{eq.ratio}, with no adjustable parameter.}
\label{fig.ratio}
\end{figure}

In Fig. \ref{fig.ratio}, we plot the measured ratio $\vdrift / \vdiff$ of the initial writing speeds for the diffusive mechanism (with $E_0 =0$), and the drift mechanism (with $E_0 = 2$kV/cm), as function of the lattice period $d$. The ratio $\vdrift / \vdiff$ increases linearly as expected with the lattice constant, and the measured slope is very close to the theoretical expectation from Eq. \ref{eq.ratio}, with $T=300$K (solid line), which confirms our analysis and validates Eq. \ref{eq.v0}.

\section{IV. Observations of linear and non-linear wave propagation in regular lattices}

In this section we study patterns of propagation in simple lattices, for plane waves and gaussian wave packets.

\subsection{1. Linear wave propagation}

\begin{figure*}
\includegraphics[width=18cm]{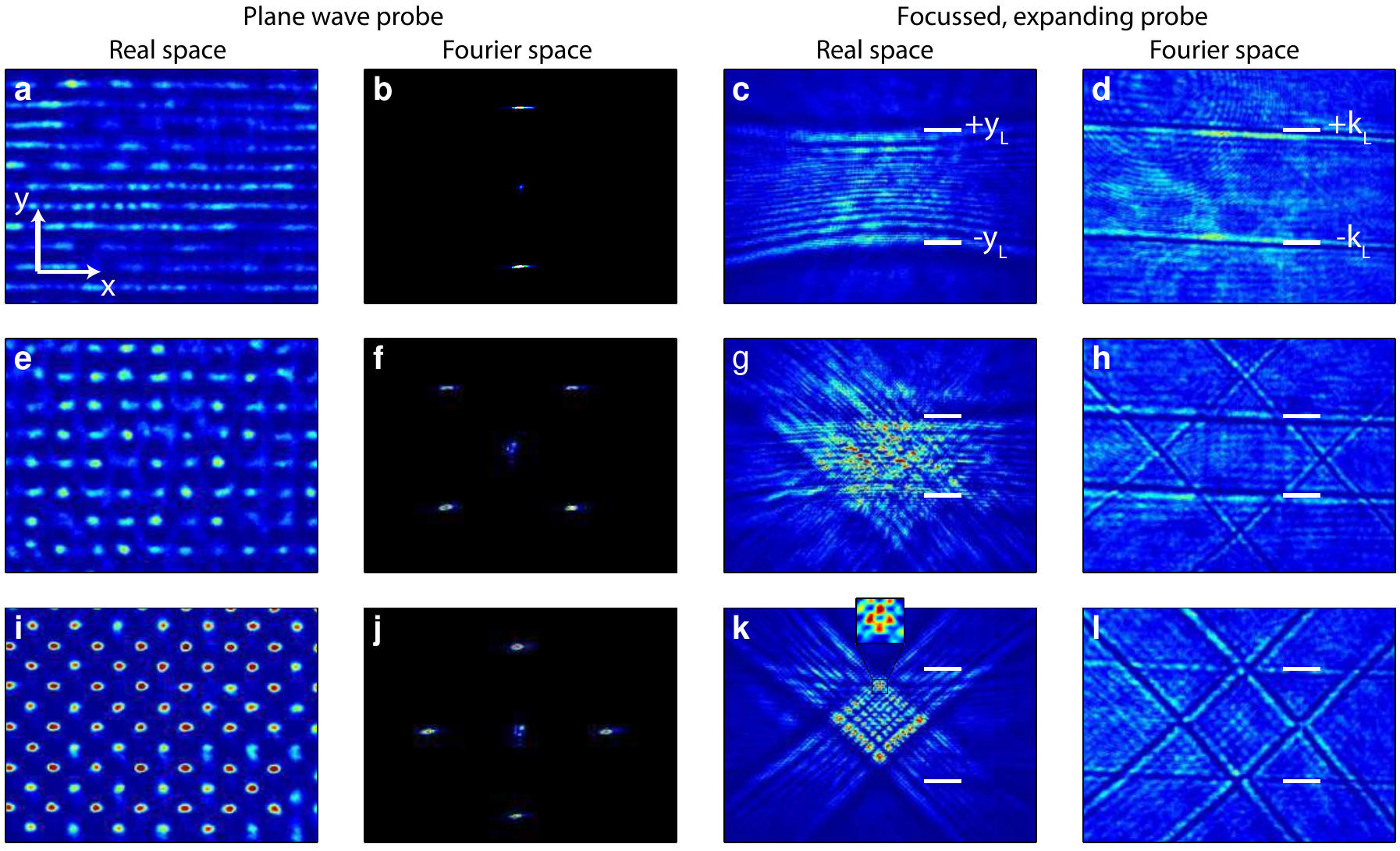}
\caption{Intensity of a linear probe beam (and writing beam, in second column) at crystal output in real and Fourier space, for a 1D (upper row), square (middle row) and a diamond lattice (lower row). The probe beam is a plane wave (first two columns) or a narrowly focused wave at the crystal input, which expands in the lattice (last two columns). 
(a,e,i) Real space output for a plane wave input probe with lattice period $d=27\mu$m (1D lattice) and $d=38.5\mu$m (2D lattices)
(b,f,j) Fourier images of lattice writing beams (four outside points) and the probe which is a point at $\kvec=0$. 
(c,g,k) Real space output for a focused probe (discrete diffraction patterns) with $d=7\mu$m (1D lattice) and $d=10\mu$m (2D lattices).
White lines show the ballistic positions $\pm \yL$ of vertical Bragg components $\pm \kL$. 
(d,h,l) Fourier images of the focused probe (Brillouin zone spectroscopy), with $d=13.6\mu$m (1D lattice) and $d=19.2\mu$m (2D lattices), with vertical Bragg components $\pm \kL$ shown as white lines.}
\label{fig.2dlatts}
\end{figure*}

Figure \ref{fig.2dlatts} shows pictures of linear wave propagation in real and Fourier space, for 1D (upper row), square (middle row), and diamond lattices (lower row). The probe beam is a plane wave (first two columns) or a beam with broad transverse spectrum, narrowly focused at the crystal input face (to a waist $w_0 =2.0 \mu$m), that expands in the crystal (last two columns). 
For the plane waves, we check in Fourier space (Fig. \ref{fig.2dlatts}.b, f, j) that the probe beam is at the center of the first Brillouin zone
\footnote{This check is important since the beam propagation in the lattice strongly depends on the input angle.}. Note that lattice periods are not the same for all types of pictures.

\subsubsection{Plane wave probe}

The real space pictures for the plane wave (Fig. \ref{fig.2dlatts}.a, e, i) correspond to a waveguiding structure analysis \cite{terhalle06}, which we use for calibrating the lattice strength \cite{Armijo14}.
Here the anisotropy of the photorefractive effect is very clear. For the square lattice, the modulation of the probe is much stronger in the vertical direction (c-axis) than in the horizontal one, as also reported in \cite{terhalle07}. 
Also, for the diamond lattice, the probe intensity at the waveguide positions is much higher than for the square lattice (although writing parameters are identical).
In our data, the anisotropy in the amplitude of refractive index modulation, is typically a factor 2, as estimated in \cite{Armijo14}.

In Fig. \ref{fig.2dlatts}.a, e, i, some imperfections are also apparent, probably attributable to residual non-linear effects, i.e., some modulational instability as discussed below. Such imperfections are particularly evident for the 1D lattice, but also for the square and diamond lattice.
In the 2D images, the irregularities could also be due to the contribution of the diffusive photorefractive effect.

\subsubsection{Focused, expanding probe wave}
\begin{figure}
\includegraphics[width=8.5cm]{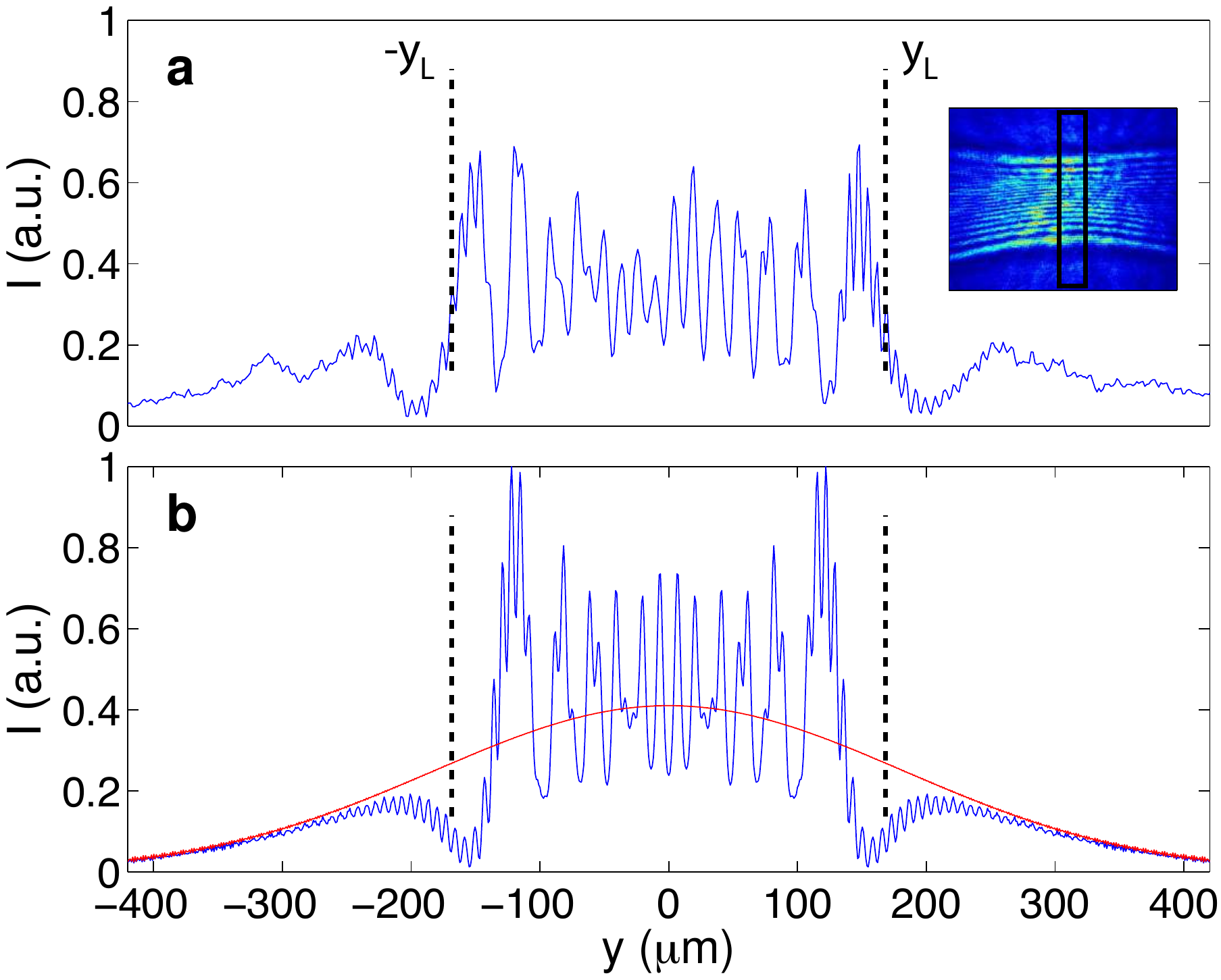}
\caption{Discrete diffraction in a 1D lattice with period $d=6.8\mu$m of an expanding wavepacket with initial waist $w_0=2.0\mu$m (data of Fig. \ref{fig.2dlatts}.c). (a) Measured profile at crystal output face (integrated in the window shown in black in inset). (b) Simulated profile in 1D lattice with lattice strength $\Delta n_0 = 0.95\times 10^{-4}$. The red solid line shows the propagation without lattice.
Vertical dashed lines show the ballistic positions $\pm \yL$ of Bragg components $\pm \kL$. The total intensity is equal in (a) and (b).}
\label{fig.dd1d}
\end{figure}

With a focused, expanding linear probe, in real space (Fig. \ref{fig.2dlatts}.c, g, k), we observe the patterns commonly called "discrete diffraction" \cite{Lederer08}, displaying two outer expanding lobes of high intensity, particularly well seen in the 1D case (Fig. \ref{fig.2dlatts}.c). 
In Fig. \ref{fig.dd1d}.a we show a vertical slice of intensity through Fig. \ref{fig.2dlatts}.c, and in Fig. \ref{fig.dd1d}.b a corresponding numerical simulation with a beam propagation code, carried with no adjustable parameters, using a sinusoidal 1D lattice whose strength $\Delta n_0= 0.95\times10^{-4}$ was determined using our calibration method \cite{Armijo14}. 
To our knowledge, previous works did not present quantitative comparisons of simulations with the measured data.
The agreement between the measured and simulated profiles is quite good, which validates our lattice calibration method.

As seen in Fig. \ref{fig.dd1d}, the outer lobes typically involve 3-4 lattice sites.
Just beyond those lobes, dark notches (or lines) are present, slightly bended in Fig. \ref{fig.2dlatts}.c due to imaging aberrations. 
Their positions correspond very well to the positions (marked as white lines in Fig. \ref{fig.2dlatts}.c,g,k, and black vertical lines in Fig. \ref{fig.dd1d}) of the ballistic propagation of wave components $\pm \kL$ (at Bragg angles)
\begin{equation}
\pm \yL = \pm \frac{\kL}{\kc} L = \pm \frac{\lambda L}{2 n_0 d},
\end{equation}
where $\kL=\pi/d$ and $\kc=2\pi \nex/\lambda$ is the wave vector modulus in the crystal.
However one can note that the measured profile is globally wider, which may be due to an imperfect matching of the focal spot of the probe beam at the crystal input face.

For the 2D square lattice (Fig. \ref{fig.2dlatts}.g), the discrete diffraction pattern features a horizontal stripe and additionally two wider diagonal stripes. 
Due to the lattice waves orientation, the horizontal stripe is narrower than in the 1D case and its edges coincides with the ballistic positions of components $\pm \sqrt{2} \kL$ (white lines).
No horizontal modulation is visible, due to the photorefractive anisotropy that causes weaker modulation of refractive index in direction $x$.
For the 2D diamond lattice (Fig. \ref{fig.2dlatts}.k), the four outer lobes form the contour of a central diamond, behind which a horizontal stripe is visible, having the same width as in the 1D case (white lines). The central zone features a well regular checkerboard pattern, with higher intensity in the four corners, each of them presenting four sites with high intensity (see inset).

For the three lattices, and especially the diamond lattice, one notices that light intensity is still present beyond the ballistic Bragg lines $\pm y_L$ (see also Fig. \ref{fig.dd1d}). This is due to the continuous character of the system.
Indeed, for a purely discrete system -the typical model being the discrete nonlinear Schr\"odinger equation (DNLSE) \cite{Lederer08}- the discrete diffraction pattern for an initial condition localized at one single lattice site is very similar to our observation, but no intensity at all is present beyond the ballistic Bragg lines (see, e.g., Fig. 1.2 or Fig. 2.7 in \cite{Lederer08}).
The intensity beyond the Bragg lines results from the spectral content of the probe beam beyond the first Brillouin zone (BZ), in the second and higher bands. Such an observation of band structure in real space has also been studied in \cite{mandelik03}. 

\begin{figure}
\includegraphics[width=8.5cm]{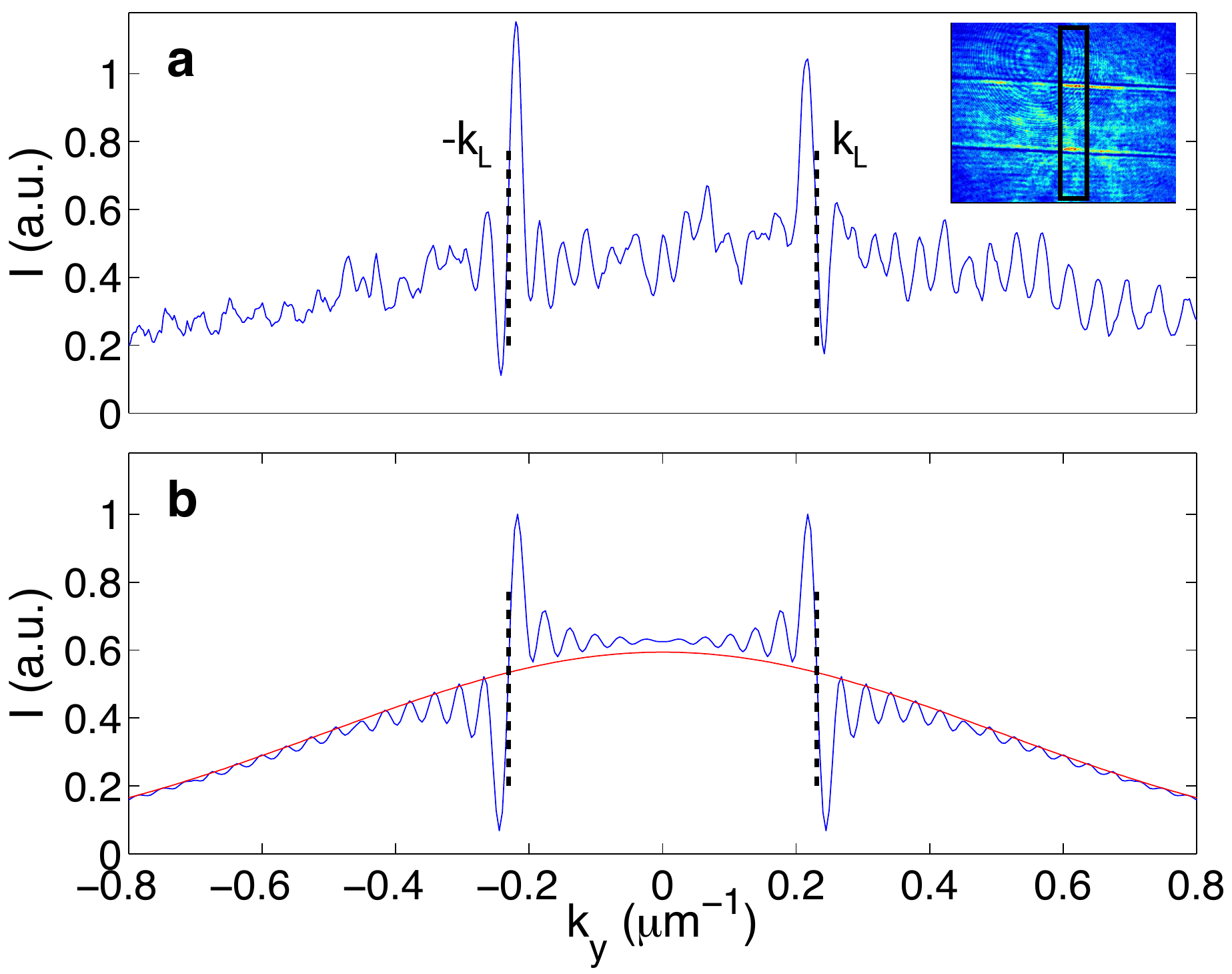}
\caption{Output intensity in Fourier space (Brillouin zone spectroscopy) for a wavepacket with initial waist $w_0=2.0\mu$m expanding in a 1D lattice of period $d=13.6\mu$m (data of Fig. \ref{fig.2dlatts}.d). (a) Measured Fourier profile (integrated in the window shown in black in inset). (b) Corresponding simulation for a 1D lattice with lattice strength $\Delta n_0 = 0.30\times 10^{-4}$. The red solid line shows the propagation without lattice.
Vertical dashed lines show the Bragg planes $\pm \kL$. The total intensity is equal in (a) and (b).}
\label{fig.bz1d}
\end{figure}

The Fourier images with focused probe (Fig. \ref{fig.2dlatts}.d, h, l) are generally referred to as "Brillouin zone spectroscopy" \cite{bartal05, terhalle06}. Interestingly, as also noted from simulations in \cite{liu10}, we obtained these pictures using a coherent probe beam (without using a spatial light diffuser as is done in \cite{bartal05, terhalle06}), and the Bragg lines are still very clearly apparent as dark notches.
The lines closest to the center mark the edge of the first BZ, matching very well the predictions $\pm \kL$ (and $\pm \sqrt{2} \kL$ for the square lattice), shown as white lines, only in the $y$ direction.
For the 2D lattices (square and diamond), higher order Bragg lines are also visible, but the anisotropy causes an almost complete absence of Bragg lines in the vertical direction.

\begin{figure*}[htbp]
\includegraphics[width=16.5cm]{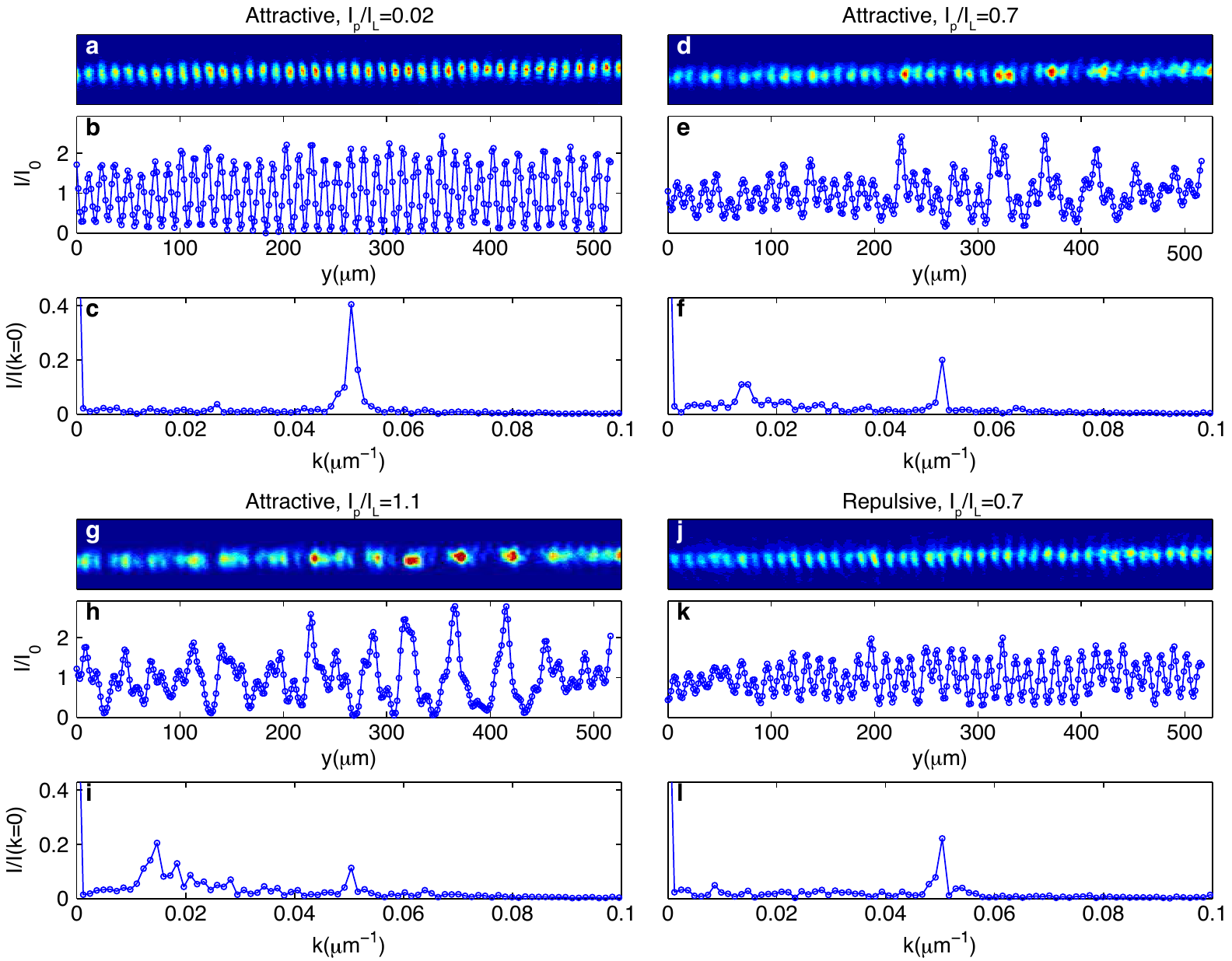}
\caption{Observation of modulational instability of a quasi-1D plane wave probe beam in a 1D lattice with period $d=10\mu$m, at $\tW=120$s, with $E_0=2$kV/cm, lattice beam intensity $\IL=0.2$mW/cm$^2$ and probe beam intensity $\Ip$. For each case the top panel shows the output intensity distribution, the middle panel, the vertically integrated intensity profile, the third panel its Fourier transform. The lattice component lies at $k\simeq 0.05/\mu$m while the MI develops at lower momenta around $k=0.015/\mu$m.
(a-c) Linear output with focusing nonlinearity and $\Ip/\IL=0.02$.
(d-f) focusing nonlinearity and $\Ip/\IL=0.7$.
(g-i) focusing nonlinearity with  $\Ip/\IL=1.1$.
(j-l) Defocusing nonlinearity and $\Ip/\IL=0.7$.}
\label{fig.MI}
\end{figure*}

In Fig. \ref{fig.bz1d}, we show a vertical profile (integrated over the window shown in the inset) for the 1D pattern of Fig. \ref{fig.2dlatts}.d, comparing it to a simulation with no adjustable parameter, using a sinusoidal 1D lattice, whose strength $\Delta n_0 = 0.30\times 10^{-4}$ was determined with our method \cite{Armijo14}. Here, as for the discrete diffraction patterns (Fig. \ref{fig.dd1d}), the simulation and measured data agree quite well, in particular, displaying two notches and two neighboring intensity maxima on the sides of the Bragg planes $\pm \kL$.
In the simulation, the oscillations decay fast away from the Bragg planes $\pm \kL$. In the experimental image, fine observation of the oscillations beyond the first cycle is rendered difficult by parasitic fringes.
We note that -to our knowledge- no analytical explanation of the Brillouin spectroscopy patterns has been proposed so far.

Overall, an important structural similarity is apparent between the discrete diffraction (Fig. \ref{fig.2dlatts}.c, g, k) and the  BZ spectroscopy pictures (Fig. \ref{fig.2dlatts}.d, h, l).

\subsection{2. Non-linear wave propagation}

For higher probe beam intensities, we observe basic nonlinear propagation phenomena.

\subsubsection{Modulational instability in quasi-1D}

Modulational instability (MI) is a general phenomenon by which an unmodulated carrier wave gets destabilized, and other frequency components grow exponentially from perturbations or background noise. MI can be considered as a precursor of soliton formation \cite{daumont97}.
In photonics, MI has been studied in many configurations including continuous photorefractive systems \cite{kip00, apolinar02, saffman04, chen02, jeng09}.
In lattice geometries, discrete MI has been observed in semiconducting waveguide arrays with a focusing nonlinearity \cite{meier04}, in the normal diffraction region (central half of the BZ), but also with defocusing nonlinearity in \cite{stepic06}, where the destabilization occurs only for carrier wave vectors lying in the region of anomalous diffraction region (outer half of the BZ, close to the band edge).

In this work we report an observation in a new quasi-one dimensional photorefractive configuration (although the physics is still governed by the full 2D and anisotropic equations Eq. \ref{eq.phi} and \ref{eq.anis}). We use a cylindrical lens of focal length $f=150$mm to focus the probe beam in only one direction, while a 1D lattice, of period $d=10\mu$m, covers the entire region of interest (in 2D).
The probe beam focus is placed at half of the crystal length $L=10$mm. Its waist in the $x$ direction is $w_0\simeq 25\mu$m and the Rayleigh length is $\lR\simeq4$mm, so that the diffraction in the $x$ direction during propagation in the crystal is unimportant.
We apply the probe and lattice writing beams together during the same writing time $\tW=120$s. (one could envision different times for both beams but this would add parameters and complexify the problem). 

In Fig \ref{fig.MI}, we show the observed output intensity at $\tW=120$s, the vertically integrated profiles $I(y)$ and their Fourier transforms $I(k)$ for the linear case (a-c), the nonlinear case with focusing nonlinearity (b-d, g-i) and for a defocusing nonlinearity (j-l).
In all cases the probe beam is launched at the center of the BZ (momentum $k=0$), and we quantify the strength of nonlinear effects by the ratio $\Ip/\IL$ of the peak probe wave intensity (at input) $\Ip$ to the average lattice beam intensity $\IL$.
In all three nonlinear cases, the writing of the lattice is fast, with an exponential time of about 30s, whereas the effects of MI develop much slower, increasing continuously over more than 200s.

In the linear case (Fig. \ref{fig.MI}.a-c, with $\Ip/\IL=0.02$), the probe beam gets modulated essentially at the lattice frequency $k\simeq0.05/\mu$m.
In the nonlinear case with moderate nonlinearity (Fig. \ref{fig.MI}.d-f, with $\Ip/\IL=0.7$), one notices the reduction of the main modulation, and the appearance of spectral weight around $k=0.015/\mu$m.
For stronger nonlinearity, (Fig. \ref{fig.MI}.g-i, with $\Ip/\IL=1.1$), the destabilization is stronger, the lattice component at $k\simeq0.05/\mu$m is almost erased and the spectral weight in the window $k=0.015-0.03/\mu$m is more important. 
This is a particularity of the photrefractive effect, where lattice writing and nonlinear effect rely on the same physical mechanism (see Section II), so that strong nonlinearity can simply erase the underlying lattice structure as in Fig. \ref{fig.MI}.g, in the zones where self-focusing is strong.

By contrast, in the defocusing case  (Fig. \ref{fig.MI}.j-l, with $\Ip/\IL=0.7$), the lattice component is almost unaffected by the nonlinearity (in the linear case, it reaches about the same value, always smaller than in the focusing case), and almost no spectral weight is apparent in the MI region, as expected for a carrier plane wave at the center of the BZ \cite{kivshar93}.

\begin{figure*}[htbp]
\includegraphics[width=13cm]{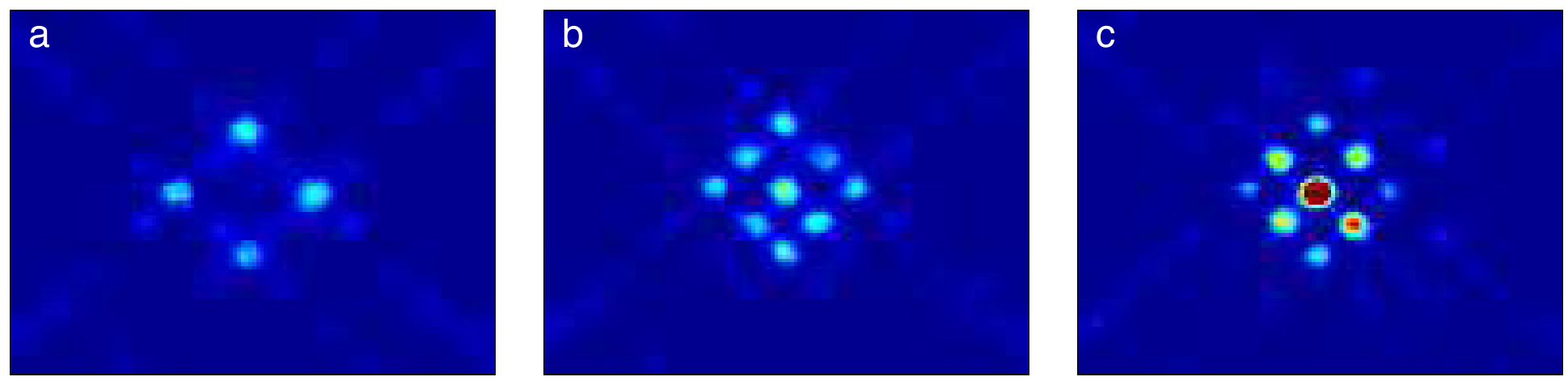}
\caption{Temporal transition from discrete diffraction to a discrete soliton in a diamond lattice of period $d= 19\mu$m with $\IL=0.8$mW/cm$^2$, at writing times $\tW=$23s (a), 39s (b), 54s (c), with same color scale. In (c), the maximal intensity in the central lobe is $\Is/\IL\simeq0.5$.}
\label{fig.DS}
\end{figure*}

\subsubsection{Temporal formation of a discrete soliton}

Although the formation of discrete solitons in diamond 2D lattices has already been reported \cite{fleischer03b, martin03}, we here present a different observation, in the time domain.
In Fig. \ref{fig.DS}, we show the output intensity at times $\tW= 23, 39, 54$s, when a diamond 2D lattice and a focused probe beam are simultaneously applied, with a ratio of peak input probe intensity to average lattice intensity $\Ip/\IL\sim 0.5$.
At short times (a), the lattice writing effect is already strong so that discrete instead of continuous diffraction is observed, with four well-marked intense outer lobes, whereas nonlinear effects have not yet noticeably come into play.
At intermediate (b) and longer times (c), the self-focusing nonlinear term leads to the formation of a discrete soliton-like propagation.
The maximal soliton intensity in units of the average lattice intensity, in (c) is $\Ip/\IL\simeq0.5$. 
In our observation, one notes the differential writing times for the lattice or nonlinear structure, as already noted in our MI measurements, which points at the complexity of the photorefractive dynamics.


\section{V. Conclusion}

We have studied basic features of the writing and probing of photorefractive lattices, providing some new verifications and observations in simple configurations.
Using linear plane waves and 1D lattices, we analyzed the often overlooked transient regime of photorefractive writing.
We checked first that the initial writing speed $v_0$ is proportional to the writing beam intensity $\IW$, as expected at high saturation $\IW \gg \Isat$.
Using $v_0$, we measured the ratio between contributions of the diffusion and drift photorefractive mechanisms, finding very good quantitative agreement with a theory assuming $v_0$ proportional to the stationary lattice strength.
We thereby provided a quantitative check of the commonly used approximation consisting in neglecting the diffusion term.

Further, we studied the linear wave propagation in regular lattices. Using plane waves, we observed effects of the photorefractive anisotropy and parasitic nonlinearities.
With focused, expanding wavepackets with a broad transverse spectrum, we analyzed discrete diffraction patterns (at finite propagation time, in real space), noting an analogy with the Fourier space patterns of Brillouin zone spectroscopy, since Bragg planes and band structure appear in both types of images.
Using our new lattice calibration method \cite{Armijo14}, for the first time to our knowledge, we could quantitatively compare experimental data to simulations based only on direct calibrations and no adjustable parameters, finding good agreement.
In the future this approach for quantitative comparisons with simulations can be extended to experiments in more complex lattices.

For nonlinear waves, we observed the development of modulational instability of plane waves in a quasi-1D geometry with focusing nonlinearity, but for a defocusing nonlinearity, no MI was observed, as expected for a carrier plane wave at the center of the Brillouin zone.
For a focused input wave, we recorded the temporal formation of discrete solitons.

In general, our work improves the understanding of photorefractive lattice experiments. Our observations of non-ideal effects (anisotropy, parasitic nonlinearity, diffusive term, non-stationarity), can explain some imperfections typical in experiments, for example the strong damping of longitudinal Talbot-like oscillations for plane waves, reported in \cite{Armijo14}, or the guided wave distortions apparent in Fig. \ref{fig.2dlatts}.a,e,j or in \cite{desyatnikov06}.

\
\begin{acknowledgments}
We aknowledge invaluable technical help from Leonardo Campos, and stimulating discussions with Martin Boguslawski, Anton Desyatnikov and Mario Molina.
Work supported by CONICYT fellowships, Programa de Financiamiento Basal de
CONICYT (Grant FB0824/2008) and Pograma ICM (Grant P10-030-F).

\end{acknowledgments}

\bibliographystyle{prsty}


\end{document}